\magnification=1100

\hsize 17truecm
\vsize 23truecm

\font\twelvec=msbm10 at 10pt
\font\sevenc=msbm10 at 7pt
\font\fivec=msbm10 at 5pt

\newfam\co
\textfont\co=\twelvec
\scriptfont\co=\sevenc
\scriptscriptfont\co=\fivec

\def\char{\mathop{\rm Char}\nolimits}

\def\Const{\mathop{\rm Const.}\nolimits}
\def\det{\mathop{\rm det}\nolimits}
\def\exp{\mathop{\rm exp}\nolimits}
\def\Exp{\mathop{\rm Exp}\nolimits}

\def\lim{\mathop{\rm lim}\nolimits}

\def\supp{\mathop{\rm supp}\nolimits}

\def\sgn{\mathop{\rm sgn}\nolimits}

\def\WF{\mathop{\rm WF}\nolimits}

\def\e{\mathop{\rm \varepsilon}\nolimits}

\baselineskip 15pt

\centerline{\bf SEMI-CLASSICAL GREEN FUNCTIONS} 
\bigskip
\centerline{A.ANIKIN${}^{1}$, S.DOBROKHOTOV${}^{1}$, V.NAZAIKINSKII${}^{1}$ {\it \&} M.ROULEUX${}^{2}$}
\bigskip
${}^{1}$ Ishlinski Institute for Problems of Mechanics
Moscow Institute of Physics and Technology, Moscow, Russia~; anikin83@inbox.ru~; dobr@ipmnet.ru~; nazay@ipmnet.ru
\smallskip 
${}^{2}$ Aix Marseille Univ, Universit\'e de Toulon, CNRS, CPT, Marseille, France~; rouleux@univ-tln.fr 
\bigskip             
\noindent {\it Abstract}: Let $H(x,p)\sim H_0(x,p)+hH_1(x,p)+\cdots$ be a semi-classical Hamiltonian on $T^*{\bf R}^n$, and 
$\Sigma_E=\{H_0(x,p)=E\}$ a non critical energy surface. Consider $f_h$ a
semi-classical distribution (the ``source'') microlocalized on a Lagrangian manifold $\Lambda$
which intersects cleanly the flow-out $\Lambda_+$ of the Hamilton vector field $X_{H_0}$ in $\Sigma_E$. 
Using Maslov canonical operator, we look for a semi-classical distribution $u_h$ 
satisfying the limiting absorption principle and 
$H^w(x,hD_x)u_h=f_h$ (semi-classical Green function).  
In this report, we elaborate (still at an early stage)
on some results announced in [AnDoNaRo1] and provide some examples, in particular from the theory of wave beams. 
\medskip
\noindent {\bf 1. Introduction}
\medskip
Our general motivation is to solve 
equations like $(H-E)u_h=f_h$, with $f_h(x)=h^{-n}f\bigl({x\over h}\bigr)$, $f\in L^2({\bf R}^n)$,
a ``localized  function'' at $x_0=0$. In case $H=-h^2\Delta$, we 
apply $h$-Fourier transform 
$${\cal F}_h u_h(\xi)=\int e^{-ix\xi/h}u_h(x)\,dx$$
to get 
${\cal F}_h u_h(\xi)={{\cal F}_1f(\xi)\over\xi^2-E}$. If $E<0$, this is an elliptic equation, and $u_h$
has the same form as $f_h$.
If $E>0$ however, 
$$u_+(x;E,h)=(2\pi h)^{-n}\int e^{ix\xi/h}{{\cal F}_1f(\xi)\over\xi^2-E-i0} \,d\xi\leqno(1)$$ 
is defined through regularisation, so to 
satisfy the limiting absorption principle. Actually, 
$$u_+(x;E,h)=E_+f_h, \quad E_+={i\over h}\int_0^\infty e^{-it(H-E)/h}\,dt$$ 
is the forward parametrix.
When $f_h(x)$ is replaced by $\delta(x-x_0)$, $u_+$ is the Green function.
Such a distribution is very singular since there are infinitely many classical trajectories linking the ``source'' $x_0$ to $x$
in time $t$; so it is suitable to consider instead ${\cal F}_1f$ with compact support. 

This problem has of course received considerable attention [Ke], [Ba], [Ku]; our special purpose is to reformulate it in terms of 
Maslov canonical operator, and describe the solution globally, including unfolding of Lagrangian singularities;
this is of special importance in the context of wave propagation. 

More generally, let $f_h=f(\cdot;h)$ be a semi-classical Lagrangian distribution (or oscillatory integral) defined on
the manifold $M$ (for all practical purposes, we shall assume here $M={\bf R}^n$), i.e. locally
$$f(x;h)=(2\pi h)^{-n/2}\int e^{i\varphi(x,\theta)/h}a(x,\theta;h)\,d\theta$$
where $\varphi(x,\theta)$ is a non-degenerate phase function in the sense of [H\"o], and 
$a(x,\theta;h)=a_0(x,\theta)+ha_1(x,\theta)+\cdots$ an amplitude. 
With $f$ we associate the critical set $C_\varphi=\{(x,\theta)\in M\times{\bf R}^N: \partial_\theta\varphi=0\}$ and 
$\iota_\varphi:C_\varphi\to T^*M$ 
with image the Lagrangian submanifold
$\Lambda_\varphi=\{(x,\partial_x\varphi): (x,\theta)\in C_\varphi\}$. Choose local coordinates $\xi\in{\bf R}^d$ on $\Lambda_\varphi$ 
and define the half-density in the local chart $(C_\varphi,\iota_\varphi)$ by
$\sqrt{d\mu_\varphi}=|\det \varphi''|^{-1/2}|d\xi|^{1/2}$. The (oscillating) principal symbol of $f$ in $\Lambda_\varphi$
is then defined (up to the factor $\sqrt{d\mu_\varphi}$) as 
$$e^{i\phi(\xi)/h}A_0(\xi)=e^{i\phi(\xi)/h}e^{i\pi\sgn\varphi''/4}a_0(x(\xi),\theta(\xi))$$
where $\phi$ is a ``reduced phase function''. 
Conversely, 
assume $\iota:\Lambda\to T^*M$ is a smooth Lagrangian immersion, with a smooth positive half-density $\sqrt{d\mu}$,
it can be parametrized locally by phase functions $\varphi$ in canonical charts $\iota_\varphi:U_\varphi\to T^*M$.
These phases can be chosen coherently, and define a class of ``reduced phase functions'' $\phi$, parametrizing $\iota$ locally.
This gives the fibre bundle of phases ${\cal L}_h$, 
including Maslov indices, equipped with transition functions.
We are also given local
smooth half-densities $|d\mu_\varphi|^{1/2}$ on $\Lambda$, defining the fibre bundle of half-densities $\Omega^{1/2}$,
equipped with transition functions. 
The collection of these objects make a fibre bundle  $\Omega^{1/2}\otimes{\cal L}_h$  over $\Lambda$. 
A section of $\Omega^{1/2}\otimes {\cal L}_h$ will be written as 
$$f(x;h)=\bigl[K_{(\Lambda,\mu)}^hA\bigr](x;h)$$
where $K_{(\Lambda,\mu)}^h$ is called Maslov canonical operator.
The ``reduced phase function'' and the ``principal symbol'' of $f$ are defined invariantly. See [M], [Du],  
[Iv], [DoZh], [DNS] for details.

So our general aim is to construct, in term of Maslov canonical operators, a representation of $E_+f$.
Here are some examples of $f$ (expressed in a single chart):

(1) $\Lambda=\{(x,\partial_x\varphi): x\in M\}$. WKB functions $f(x,h)=e^{i\varphi(x)/h}a(x,h)$ or in Fourier 
representation 
$$f(x;h)={e^{i\pi n/4}\over(2\pi h)^{n/2}}\int e^{i(xp+S(p))/h}A(p;h)\,dp\leqno(2)$$
(such an integral conveniently normalized including a phase factor, 
will be written $\int^*(\cdots)$; here $\sgn\bigl(x\cdot p)''=-n$).

(2) Gaussian functions $f(x;h)={1\over h^n}\exp(-{\omega^2\cdot x^2\over2h})$ or more general (superposition of) coherent states.

(3) A conormal distribution with $\Lambda=T^*_N{\bf R}^n$, $N=\{x_n=0\}$
$$f(x;h)=\int^* e^{ix_np_n/h}A(x',p_n)\,dp_n\leqno(3)$$

(4) $\Lambda=\{x=X(\varphi,\psi)=\varphi\omega(\psi), p=P(\varphi,\psi)=\omega(\psi), \varphi\in{\bf R}\}$, 
$\omega\in {\bf S}^{n-1}$, $f$ identifies with a ``Bessel beam'', see Sect. 4.
\bigskip
\noindent{\bf 2. Hypotheses and main result}
\medskip
Let $\Lambda_0\to T^*M$, $\Lambda_1\to T^*M$ be smooth embedded Lagrangian manifolds, $\Lambda_1$ 
with a smooth boundary $\partial\Lambda_1$ (isotropic manifold). Following [MelUh], we say that 
$(\Lambda_0,\Lambda_1)$ 
is an {\it intersecting pair} iff $\Lambda_0\cap\Lambda_1=\partial\Lambda_1$ and the intersection is {\it clean}, i.e.
$$\forall \rho\in\partial\Lambda_1 \quad T_\rho\Lambda_0\cap T_\rho\Lambda_1=T_\rho\partial\Lambda_1$$
In particular, near each $\rho\in T^*M$, there is a canonical transform mapping $\Lambda_0$ to $T_0^*{\bf R}^n=\{x=0\}$ 
and $\Lambda_1$ to 
$$\Lambda_+^0=\{(x,\xi): x=(0,x_n), \xi=(\xi',0), x_n\geq0\}$$
i.e. the flow-out of $T_0^*{\bf R}^n$ by the Hamilton vector field
$X_{\xi_n}=((0,\cdots,0,1),0)$ of $\xi_n$, passing through $x=0, \xi_0=(\xi',0)$. 

Let now $H(x,p;h)=H_0(x,p)+h H_1(x,p)+\cdots$ be a symbol of class $S^0(m)$ 
and $H^w(x,hD_x;h)$ its Weyl quantization, as a bounded operator on $L^2(M)$. Assume 
\smallskip
\noindent (A.1) The energy surface $\Sigma_E=H_0^{-1}(E)$ is  non critical.
\smallskip
\noindent (A.2) The set $L=\Sigma_E\cap\Lambda$ is compact.
\smallskip
\noindent (A.3) The Hamiltonian vector field $X_{H_0}$ is transverse to $\Lambda$ along $L$.
\smallskip
For $t\geq0$, let $g_{H_0}^t=\exp tX_{H_0}$, $\Lambda_t=g_{H_0}^t(\Lambda)$, and 
$\Lambda_+=\bigcup_{t\geq0}g_{H_0}^t(L)$.
The pair $(\Lambda,\Lambda_+)$ define a Lagrangian intersection, and $L=\partial\Lambda_+$. 
We assume also 
\smallskip
\noindent (A.4) The Hamilton flow $g_{H_0}^t|_L$ extends for all $t>0$. 
\smallskip
\noindent (A.5) $|\pi_x(g_{H_0}^t|_L)|\to\infty$ as $t\to\infty$ (non trapping condition). 
\smallskip
Let $\psi$ be a system of coordinates on $L$, completed (locally near $L$) to coordinates $(\psi,\tau)$ on $\Lambda$ 
and to coordinates $(\psi,t)$ on $\Lambda_+$,
so that the positive measure on $\Lambda$ (square of half-density) takes the form $d\mu=|d\tau\wedge d\psi|$. This implies the measures
$d\mu_t=(g_{H_0}^{t*})^{-1}(d\mu)$ on $\Lambda_t$ and 
$d\mu_+=|dt\wedge\,d\psi|$ on $\Lambda_+$. 

We shall consider a partition of unity subordinated to a covering of the pair $(\Lambda,\Lambda_+)$, with cut-off functions
(of small enough support) $\widetilde\chi_0\in C_0^\infty(\Lambda)$, that depends only on $\tau$ 
(the ``radial variable'' on $\Lambda$), with $\widetilde\chi_0(\tau)\equiv 1$ near $L=\{\tau=0\}$, 
and $\chi_0\in C_0^\infty(\Lambda_+)$, that depends only on $t$, $\chi_0(t)\equiv1$ near $t=0$. Let also
$\chi_T\in C^\infty({\bf R}^+)$, $\chi_T\equiv1$ for  $t\leq T/2$, $\chi_T\equiv0$ for  $t\geq T/2$
where $T$ is so large that $x\notin\pi_x(g_{H_0}^t|_L)$ if $t\geq T$
(using the non-trapping condition (A.5)). 

For simplicity, we state our main result [AnDoNaRo1]
in the case $\Lambda=T_{0}^*M$, so that all solutions of Hamilton 
equations start from
some $(0,p)$, with $p\in \supp A$. 
Representing $E_+$ as a sum of Maslov canonical operators associated with the pair $(\Lambda,\Lambda_+)$
in particular we shall retrieve the theorem on propagation of singularities for wave-front sets
$\WF _h u\subset \WF _h (Hu)\cup \char H$.
\medskip
\noindent{\bf Theorem 1}: {\it Let $\Omega\subset M$ be a bounded domain.
Under hypotheses (A.1)-(A.5) above, there is $u=u_h$, solution of $H^w(x,hD_x;h)u_h=f_h$,
satisfying the limiting absorption principle, which
can generically be cast in the following form
$$\eqalign{
&u(x,h)=\int^*{(1-\widetilde\chi_0(\tau(p))A(p)\over H(0,p)} e^{ipx/h}\,dp \cr
&+\int_0^{\e _0}\,dt\int^* e^{i\Phi(t,x,\eta)/h}e^{-i\Theta(t,x,\eta)}
|\det J(t,x,\eta)|^{-1/2}
\chi_T(t)\chi_0(t)\rho\bigl(\tau(\eta)\bigr)A(\eta)\,d\eta\,dt \cr
&+K^h_{(\Lambda_+,d\mu_+)}\bigl[\chi_T(\cdot)(1-\chi_0(\cdot))e^{-i\Theta}A_+\bigr] \cr
}\leqno(4)$$
up to ${\cal O}(h)$ terms; the summands on the right hand side are boundary, transient and wave part,
respectively. In Eq.(4), $\Phi(t,x;\eta)$ solves Hamilton-Jacobi equation (see Sect.3), and
$\Theta(t,x,\eta)=\int_0^tH_1\circ\exp (sX_{H_0}(0,\eta))\,ds$ is the integral of the sub-principal 1-form,
where we recall $H_1$ is the sub-principal symbol of $H$. 
Moreover $J(t,x,\eta)$ is a Jacobian computed from Hamilton-Jacobi equation. }
\medskip
The case when $\Lambda$ is slightly ``tilted'' with respect to the vertical plane, i.e. 
has generating function $\phi(x,\xi)=x\xi+S(\xi)$
can be formulated in a very similar way (see Sect.3). 
For more general $f$'s, the boundary part should be written as $\bigl[K_{(\Lambda,\mu)}^h{(1-\widetilde\chi_0)A\over H}\bigr](x;h)$, 
and the transient part
as the integral over $t$ of a phase factor times $\bigl[K_{(\Lambda_t,\mu_t)}^h{\chi_T(\cdot)\widetilde\chi_0(\cdot)\chi_0(\cdot)A_t}\bigr](x;h)$, 
$A_t$ being the 
solution of the transport equation along the projection of $\exp tX_{H_0}$ with initial data $A(\eta)$. 

The boundary part is microlocalized on $\Lambda$, the wave part on $\Lambda_+$. Let us make first a few comments:

1) (4) can be interpreted as ``integrated'' Van Vleck Formula [CdV],
which expresses the semi-classical propagator $e^{-itH/h}$ acting on a localized function.  

2) The wave part contributes generally at $x\in\Omega$ only if 
$t\mapsto\Phi(t,x,\eta)$ 
has a non-degenerate critical point for all $\eta\in\supp f$: 
let  $hD_{x_n}$ be the ``model'' operator, and $f(x;h)=\int ^*e^{ixp/h}A(p)\,dp$. 
Then [MelUh]
$$u_h(x)={i\over h}\int_0^\infty\chi_T(t)\,dt\int ^* e^{i(x'\xi'+(x_n-t)\xi_n)/h}A(\xi)\,d\xi$$
verifies the limiting absorption principle and
$hD_{x_n}u(x,h)=f(x,h)+{\cal O}(h^\infty)$ for $x_n\leq T/2$; still $u_h$ has no ``wave part''. 

3) Formula (4) is also valid near a focal point $x\in\Omega$, 
or more generally when $x$ is linked to $x_0$ through a trajectory containing
several focal points. The wave-part simplifies outside the focal points to a WKB form,
involving non trivial Maslov indices passing the first focal point.

4) The phase function $\Phi(t,x,\eta)$ can be also replaced by a Lagrangian action, in the spirit of [DNS].
This will be discussed in detail in a future work. 
\medskip
We illustrate Theorem 1 by computing $u_h$ explicitely in the 2-D case for Helmholtz operator with 
constant coefficient as in (1), but $f$ with compact support. Let $f$ also be radially symmetric;
its Fourier transform $g={\cal F}_1f$ is again of the form 
$g(p)=g(|p|)=g(r)$ and extends holomorphically to ${\bf C}^2$. For $E=k^2$, $k>0$,
we rewrite (1) as $u_h(x)=u(x)=u_0(x)+u_1(x)$ with
$$\eqalign{
&u_0(x)={k+i\e \over(2\pi h)^2}\int_0^{2\pi}\,d\theta\int_0^\infty\exp[i|x|r\cos\theta/h]{g(r)\over r^2-(k+i\e )^2}\,dr\cr
&u_1(x)={1\over(2\pi h)^2}\int_0^{2\pi}\,d\theta\int_0^\infty\exp[i|x|r\cos\theta/h]{g(r)\over r+k+i\e }\,dr\cr
}$$
To compute $u_0$ we use contour integrals. When $\theta\in]-{\pi\over2},{\pi\over2}[$, we shift the contour of integration
to the positive imaginary axis and get by the residues formula
$$\eqalign{
\int_0^\infty&\exp[i|x|r\cos\theta/h]{g(r)\over r^2-(k+i\e )^2}\,dr+\int_0^\infty\exp[-|x|r\cos\theta/h]{g(ir)\over r^2+(k+i\e )^2}\,idr=\cr
&2i\pi{g(k+i\e )\over2(k+i\e )}\exp[i|x|(k+i\e )\cos\theta/h]
}\leqno(5)$$
while for $\theta\in]{\pi\over2},{3\pi\over2}[$,
$$\int_0^\infty\exp[i|x|r\cos\theta/h]{g(r)\over r^2-(k+i\e )^2}\,dr-\int_0^\infty\exp[|x|r\cos\theta/h]{g(-ir)\over r^2+(k+i\e )^2}\,idr=0
\leqno(6)$$
Summing up (5) and (6), integrating over $\theta\in]0,2\pi[$ and letting $\e \to0$, we obtain
$$\eqalign{
u_0(x)&={i\pi g(k)\over (2\pi h)^2}\int_{-\pi/2}^{\pi/2}\exp[i|x|k\cos\theta/h]\,d\theta+\cr
&\int_0^\infty{dr\over r^2+k^2}\bigl[\int_{-\pi/2}^{\pi/2}g(ir)-\int_{\pi/2}^{3\pi/2}g(-ir)\bigr]\exp[-|x||\cos\theta|/h]\,d\theta
}$$
Since $g(ir)=g(-ir)$, the latter integral vanishes, so we end up with
$$u_0(x)={i\pi g(k)\over (2\pi h)^2}\int_{-\pi/2}^{\pi/2}\exp[i|x|k\cos\theta/h]\,d\theta$$
It is readily seen that 
$$\WF _h u_0\subset\{x=0\}\cup\{(x,k{x\over|x|}), x\neq 0\}=\Lambda\cup\Lambda_+$$
Consider now $u_1$. We let $\e \to 0$ and set $\widetilde g(r)={g(r)\over r(r+k)}$. Since
$\widetilde g(r)\sqrt r\in L^1({\bf R}_+)$, we have 
$u_1(x)=H_0(\widetilde g)({|x|\over h})$, where $H_0$ denotes Hankel transform of order 0. 

Let $\chi\in C_0^\infty({\bf R}^2)$ be
radially symmetric, and equal to 1 near 0, since $\WF _h f_h=\{x=0\}$, we have 
$$g={\cal F}_h(\chi f_h)+{\cal O}(h^\infty)=(2\pi h)^{-2}{\cal F}_h(\chi)*g+{\cal O}(h^\infty)$$
so in the expression for $u_1$ we may replace mod ${\cal O}(h^\infty)$, $\widetilde g(r)$ by a constant times
$\widehat g(r)={({\cal F}_h(\chi)*g)(r)\over r(r+m)}$ (see [Bad] for 2-D convolution and Fourier transform in polar coordinates). 
To estimate $\WF _h u_1$, we compute again the Fourier transform of $(1-\widetilde\chi)\widehat g$ where $\widetilde \chi$
is a cut-off equal to 1 near 0, and we find it is again ${\cal O}(h^\infty)$ if $\chi\equiv1$ on $\supp\widetilde\chi$.
This shows that  
$\WF _hu_1\subset \{x=0\}$. See also 
[MelUh], Prop. 2.3. 
Note that the decomposition $u_0+u_1$ is directly related with the corresponding one in Theorem 1 as the sum of 
the boundary ($u_1$), and the wave part ($u_0$). We conjecture that the transient part in Theorem 1 can be removed
(taking a limit $\supp\chi_0\to\emptyset$) when the Hamiltonian flow
$X_{H_0}$ enjoys some non-degeneracy properties (see Sect.3), as in the case of a geodesic flow. 
\bigskip
\noindent{\bf 3. Maslov canonical operators associated with $(\Lambda,\Lambda_+)$}
\medskip
Proof of Theorem 1 consists first in looking at the propagator $e^{-itH/h}$ acting on Lagrangian distribution $f_h$ as above, i.e. the
solution of the Cauchy problem $\bigl(hD_t+H^w(x,hD_x)\bigr)v=0$, $v|_{t=0}=f_h$.
Next step is to integrate with respect to $t$ after introducing the partition of unity above.

In this report, we shall content to construct the phase functions by solving Hamilton-Jacobi equation.
First we recall from [H\"o],Thm 6.4.5 the following:
\medskip
\noindent{\bf Theorem 2}: {\it Denote the variable in $T^*{\bf R}^d$ by $(y,\eta)=(y',y_d;\eta',\eta_d)$. Let ${\cal H}$ be a real valued, 
smooth Hamiltonian 
near $(0,\eta)$ such that
${\cal H}(0,\eta)=0$, $\partial_{\eta_d} {\cal H}(0,\eta)\neq 0$,
and let $\phi$ be a real valued, smooth function on ${\bf R}^{d-1}$ such that 
$\partial_{y'} \phi=\eta'$.
Then there exists in a neighborhood of $0\in{\bf R}^d$ a unique real valued solution $\Phi(y;\eta)$ of 
Hamilton-Jacobi equation ${\cal H}(y, \partial_y\Phi)=0$
satisfying the boundary condition}
$$\Phi(y',0;\eta)=\phi(y'), \quad {\partial \Phi\over\partial y}(0;\eta)=\eta$$
\smallskip
We consider here the case of a ``maximally singular'' chart $U$ for $\Lambda$ 
where $\Lambda=\Lambda_\phi$ has generating function $\phi(x,\xi)=S(\xi)+x\xi$. Even if $\Lambda$ is a plane, 
the outgoing manifold $\Lambda_+$ may be very complicated far away from $x_0$, but changing the canonical charts,
we can proceed step by step.  

Without loss of generality we can also assume here $E=0$, $S(0)=0$, $\partial_\xi S(0)=0$. Let $\eta,\tau$ such that $\tau+H(0,\eta)=0$, by Theorem 2
(after slightly changing the notations), there exists $\Phi(x,t;\eta,\tau)$ such that 
$$\eqalign{
&{\partial \Phi\over\partial t}+H(x,{\partial\Phi\over\partial x})=0\cr
&\Phi(x,0;\eta,\tau)=S(\eta)+x\eta\cr
&{\partial \Phi\over\partial x}(0,0;\eta,\tau)=\eta, \quad {\partial \Phi\over\partial t}(0,0;\eta,\tau)=\tau\cr
}\leqno(7)$$
Moreover, when $\tau$ is small enough, the intersecting pair $(\Lambda,\Lambda_+)$ for energy level $H=0$
extends to a smooth family of intersecting pairs $(\Lambda,\Lambda_+(\tau))$ for energy levels $H=-\tau$ 
$$\Lambda_+(\tau)=\{(x,\xi)\in T^*M, \exists t\geq0, \exists(y,\eta)\in\Lambda, \ (x,\xi)=\exp tX_H(y,\eta), \tau+H(y,\eta)=0\}$$
intersecting along 
$L(\tau)$ given by $\eta_n=\widetilde\eta_n(\eta',\tau)$,
with $\widetilde \eta_n(\xi'_0,0)=0$. 

So near $x=x_0$ we can assume, 
possibly after permuting the $\xi$-coordinates, 
that $H(0;\xi'_0,0)=0$, ${\partial H\over\partial\xi_n}(0;\xi'_0,0)\neq0$,
and for $\tau$ small enough, the equation $\tau+H(x,\xi)=0$ is equivalent to $\xi_n=\widetilde \xi_n(x,\xi',\tau)$,
with $\widetilde \xi_n(0,\xi'_0,0)=0$.

Now we want to set $\tau=0$ and therefore, given $(x,\eta)$, solve the equation ${\partial \Phi\over\partial t}=0$ for $t\geq0$.
As we have seen, uniqueness of solutions doesn't always hold, as shows the ``model case''
$hD_{x_n}$. Namely, the phase function given by Hamilton-Jacobi theory is 
$\Phi(x,t;\eta)=\phi(x',x_n-t,\eta)$, so $\partial_t\Phi=0$ iff $\eta_n=0$, for all $t$.
However the phase parametrizing $\Lambda_+$
is given for small $t$ by Taylor expansion
$$\Phi(x,t;\eta)=x\eta+S(\eta)-tH(x,\eta)
+{t^2\over2}\partial_\xi H(x,\eta)\partial_x H(x,\eta)+\cdots$$
The following assumption in turn ensures existence of a finite number of 
solutions $t=t(x;\eta)>0$ of ${\partial \Phi\over\partial t}=0$.
Denote by $\Exp_{x_0}^t\eta=\pi_x(\exp tX_{H_0}(x_0,\eta))$ the projection of the bicharacteristic of $H$ starting from $(x_0,\eta)$
near $(x_0,\xi_0)$ at time 0. Assume:
\smallskip
\noindent $(A'_0)$ {\it $\exists \Omega_\xi\subset{\bf R}^n$ 
open (small) neighborhood of $\xi_0$, 
such that if $x=\Exp_{x_0}^t\eta$ for some $t>0$, and $\eta\in\Omega_\xi$, then the map
$\xi\mapsto \Exp^t_{x_0}\xi$ is a local diffeomorphism near $\eta$; in other terms,
$x_0$ and $x$ are not conjugated along any trajectory that links them together within time $t$,
with initial momentum $\xi$. }
\smallskip
The set of such $x$ is an open set $\Omega_x$. 
\medskip
\noindent{\bf Proposition 1}: {\it Under hypothesis $(A'_0)$, for all $(x,\eta)\in\Omega_x\times\Omega_\xi$
there is a finite number of $t_j=t_j(x,\eta)>0$ ($1\leq j\leq N$) solutions of 
$\partial_t\Phi(x,t;\eta)=0$ and $t_j$ are non degenerate critical points. }
\medskip
Moreover $(A'_0)$ is generically fulfilled before occurence of the first focal point, as show the 
following examples.
\smallskip
Let $H_0$ be a geodesic flow, $E>0$, $f$ conormal to $N=\{x_0\}$~; $\Omega_x$ is a small enough ball centered at $x_0$,
(geodesic ball), or $\Omega_x$ is a neighborhood of a minimal geodesic 
$\{y=\Exp_{x_0}^s\eta, 0\leq s\leq t_0\}$ for $\eta$ in some small neighborhood of $\eta_0$.
This applies when $f$ is as in (2) and $\supp A$ is localized near $\eta_0$. 
The same holds with $f$ as in (3) conormal to the hypersurface $N=\{x_n=0\}$
(or more generally to a surface $N$ of positive codimension in $M$). For instance when $N=\{x_n=0\}$,
$(A'_0)$ holds with $\Omega_\xi$ 
an open (small) neighborhood of $\eta=(0,\cdots,1)$, and $\Omega_x$ a neighborhood of a minimal geodesic from $N$ to
some $x_1$. This follows from
a well-known property of minimal geodesics (see e.g. [HeSj], Proposition 6.3). 
In case of a Schr\"odinger operator with principal symbol $H_0(x,\xi)=\xi^2+V(x)-E$ (where $E$ is a scattering energy)
we use Maupertuis-Jacobi principle to reduce again to a Riemannian metric $d_E$ given by
$ds=(E-V(x))^{1/2}|dx|$. 
So we arrive at the same conclusions, the metric $d_E$ being conformal to the standard metric. 
We proved the following:
\medskip
\noindent{\bf Proposition 2}: {\it Under the minimality assumptions above, $(A'_0)$ holds.
For $\eta\in\Omega_\xi$ and $x=Exp^t_{x_0}(\eta)\in\Omega_x$, $0<t<t_0$, the Lagrangian manifold $\Lambda_t=\exp tX_H(\Lambda_\phi)$ 
has same rank as $\Lambda$, and is of the form $\Lambda_{\Phi(t,\cdot)}$. In particular
$$\exp tX_{H_0}(\Lambda_\phi)=\{(x,\partial_x\Phi(t,x;\eta)): \partial_\eta\Phi(t,x;\eta)=0\}
\subset\{\exp tX_{H_0}(x,\eta), (x,\eta)\in T^*N^\perp\}$$
Moreover for all $(x,\eta)\in\Omega_x\times\Omega_\xi$
there is a unique $t(x,\eta)>0$ solution of 
$\partial_t\Phi(x,t;\eta)=0$ and $t(x,\eta)$ is a non degenerate crtical point. }
\medskip
Extending the geodesic $\gamma$ beyond the first focal point occuring in some $\Lambda_{t}$ for some $t=t_0$.
we only need another representation of the phase function. Using the mixed representation for Lagrangian manifolds, we know
that for any $k=0,\cdots,n$, ($k=0$ corresponds to a maximally singular chart, $k=n$ to a regular chart, they are mapped onto
each other by Fourier transform), 
there exists a partition of variables $x=(x',x'')\in{\bf R}^k\times{\bf R}^{n-k}$ and $\xi=(\xi',\xi'')$,
such that if 
$\widetilde\pi:{\bf R}^{2n}\to{\bf R}^n$, $(x,\xi)\mapsto(x',\xi'')$, then rank $d\widetilde\pi=n$. In such a chart,
the generating function for $\Lambda_t$ takes the form $\phi(x,\eta)=x''\eta''+S(x',\eta'')$. We can reformulate
Hamilton-Jacobi equations as in (7) in these coordinates, which has again a unique solution for small $t-t_0$.
Generically (i.e. under an assumption $(A'_k)$ modeled after $(A'_0)$ in the new $(x,\xi)$ 
coordinates), one still obtains a non degenerate phase function, which contributes to the fibre bundle ${\cal L}_h$
over $\Lambda_+$ together with Maslov index of $\gamma$. More explicit formulae will be given elswewhere.
Of course, $\gamma$ may contribute a finite number of times in the expression for the Green function at $x$,
and a finite number of $\gamma$ contribute to the wave part of the phase function.
\bigskip
\noindent{\bf 4. Using eikonal coordinates}.
\medskip
The computations above can be simplified using special coordinates adapted to $\Lambda$, called {\it eikonal coordinates}. 

Let $\iota:\Lambda\to T^*M$ be a smooth embedded Lagrangian manifold.
The 1-form $p\,dx$ is closed on $\Lambda$, so locally exact, and $p\,dx=dS$ on any simply connected domain $U$.
Such a $S$ is called an {\it eikonal} (or action) and is defined up to a constant. Assume $dS\neq0$ on $\Lambda$, 
then $S$  can be chosen as a coordinate on $U$, which we complete by smooth functions $\psi\in{\bf R}^{n-1}$. 

We use here eikonal coordinates to construct a phase function solving Hamilton-Jacobi equation in case of a positively
homogeneous Hamiltonian of degree $m$ with respect to $p$. To fix ideas, we take $n=2$ (for simplicity) $m=1$ and
$$H(x,p)={|p|\over n(x)}\leqno(8)$$
\medskip
\noindent {\bf Example 1}: $\Lambda=\{x=0\}$ intersects the energy
surface $H=1$ along $L$. Let us compute the eikonal $S$ on $\Lambda_+$. Integrating Hamilton equations we have
$x=X(t,\psi), p=P(t,\psi)$ where $\psi\in{\bf R}^2$, hence
$dS=p\,dx|_{\Lambda_+}=\langle P(t,\psi),dX(t,\psi)\rangle$.
Since $dx=0$ on $\Lambda$, we have $S(0,\psi)=\Const =S_0$, and 
$$S(t,\psi)=S(0,\psi)+\int_{(0,\psi)}^{(t,\psi)}p\,dx|_{\Lambda_+}=S_0+\int_0^t\langle P(s,\psi),\dot X(s,\psi)\rangle\,ds\leqno(9)$$
By Hamilton equations and Euler identity, we have on $H=1$. 
$$\langle P(s,\psi),\dot X(s,\psi)\rangle=
\langle P(s,\psi),\partial_pH(X,P\rangle=mH(X,P)=m$$
and $S(t,\psi)=S_0+mt$ is the action on $\Lambda_+$. So
$$m\,dt=\langle P(t,\psi),dX(t,\psi)\rangle=\langle P(t,\psi),\dot X(t,\psi)\rangle\,dt+\langle P(t,\psi),X_\psi(t,\psi)\rangle\,d\psi$$
It follows that $\langle P(t,\psi),\dot X(t,\psi)\rangle=m$ and $\langle P(t,\psi),\partial_\psi X(t,\psi)\rangle=0$. 
Now we complete the coordinate system $\psi$ on $\Lambda$ by a smooth function $r$ such that $L$ is given by $r=1$, and set 
$$\Phi(x,(t,\psi,r))=mt+r\langle P(t,\psi),x-X(t,\psi)\rangle$$
where $r$ can be interpreted as a Lagrange multiplier. Let us check that $\Phi$ satisfies Hamilton-Jacobi equation. We have
$$\eqalign{
&\partial_t\Phi=\dot\Phi=m+r\langle\dot P,x-X(t,\psi)\rangle-r\langle P,\dot X\rangle=\cr
&m(1-r)+r\langle\dot P(t,\psi),x-X(t,\psi)\rangle\cr
&\partial_r\Phi=\langle P(t,\psi),x-X(t,\psi)\rangle\cr
&\partial_\psi\Phi=r\langle \partial_\psi P(t,\psi),x-X(t,\psi)\rangle\cr
}\leqno(10)$$
Last 2 equations in $(\dot\Phi,\partial_r\Phi,\partial_\psi\Phi)=0$ give an homogeneous linear system with determinant $\det(P,P_\psi)$.
On $x=0$ we get $|p|=n(0)>0$, and in dimension $n=2$, $p=n(0){}^t(\cos\psi,\sin\psi)$. It follows that det $(P,P_\psi)=|n(0)|^2$ so
for small $t$, we get $x-X(t,\psi)=0$, so the phase is critical with respect to $(\psi,r)$ for $x=X(t,\psi)$. 
Substituting into the last equation (10) we get $\dot\Phi=0$ when $r=1$.
We complete the proof that $\Phi$ is a generating function for $\Lambda_+$ by checking 
$d\partial_{t}\phi,d\partial_{\psi}\Phi,d\partial_r\Phi$ are linearly independent on the set $x=X(t,\psi)$.
This can be done by examining the variational system associated with Hamilton equations. 

Moreover we can reduce this generating function by eliminating $t$ by stationary phase. Again, of course, this holds only for small $t>0$,
before unfolding of Lagrangian singularities.
\medskip
\noindent {\bf Example 2}: We take $H$ as in (8) with $n(x)=n(|x|)$, $n=2$ for simplicity and 
$$\Lambda=\{x=X(\varphi,\psi)=\varphi\omega(\psi), p=P(\varphi,\psi)=\omega(\psi),\ 
\varphi\in{\bf R}, \omega\in {\bf S}^{n-1}\}$$
$\psi$ being the usual angles parametrizing $\omega\in {\bf S}^{n-1}$. 
When $n=2$ this is the wave-front set of Bessel function $f_h(x;h)=J_0({|x|\over h})$; such functions arise 
in the wave beam theory (see [Ki], [DoMaNa] and references therein), so we call $f_h$ a ``Bessel beam''. 

Computing the action  we find
$p\,dx|_\Lambda=d\varphi$ 
so coordinate $\varphi$ will play the role of $x$ in he previous Example. We have $n(|x|)=n(\varphi)$,
and $\varphi=\Const $ on $L=\Lambda\cap\{H=1\}$. The argument above extends readily to this setting, in particular as in (9)
$$\eqalign{
S(t,&\varphi,\psi)=S(0,\varphi,\psi)+
\int_{(0,\varphi,\psi)}^{(t,\varphi,\psi)}p\,dx|_{\Lambda_+}=
S(0,\varphi,\psi)+\int_{(0,\varphi,\psi)}^{(t,\varphi,\psi)}P(s,\varphi,\psi)\,dX(s,\varphi,\psi)=\cr
&S_0+\int_0^t\langle P(s,\varphi,\psi),\dot X(s,\varphi,\psi)\rangle\,ds
}$$
and 
$$\langle P(t,\varphi,\psi),\partial_\psi X(t,\varphi,\psi)\rangle=
\langle P,\partial_\varphi X\rangle=0$$
Now we can apply the results of [DNS], Sect.2.2 on eikonal coordinates for a Lagrangian manifold in a general position, with
$X_{\widetilde\psi}(\widetilde\varphi,\widetilde\psi)$ of rank $k$. Here $\widetilde\varphi,\widetilde\psi$ are coordinates in a local
chart of $\Lambda_+$ in the extended phase space. 
More specifically, we take $\widetilde\psi=(t,\psi)$, $\widetilde\varphi=\varphi$, try to make a change of variables
$\varphi=\varphi(x,t,\psi)$, and seek for a generating function of $\Lambda_+$ in the form
$\Phi(x,(t,\psi,r))$ as in the previous Example. Details will be given elsewhere.
\bigskip
\noindent {\bf 5. More examples and perspectives}.
\medskip
The methods above apply in a number of situations as:

(1) The water-wave Hamiltonian $H(x,hD_x;h)$ with $H_0(x,p)=|p|\tanh\bigl(|p|D(x)\bigr)-E$, together with $\Lambda=\{x=x_0\}$
has been discussed in [DoNa], [AnDoNaRo2], in relationship with (11) or Helmhotz operator $H(x,hD_x)=-h^2\Delta-\bigl(n(x)\bigr)^2$.
The localized function $f$ can be a Gaussian (even in $x$) or a Gaussian times a linear function (odd in $x$),
or can be of antenna type, i.e. its Fourier transform localized in a cone in $p$. 

(2) The kinetic part of Hamiltonian is of Lorenzian type (as $-p_0^2+p_1^2$), and $f$ a localized function (Gaussian)
supported on $\Lambda=\{x=x_0\}$, so that the semi-classical Green function $u_h$ is the 
linear response to $f$ localized on Kelvin angle (or Mach cone). 
  
It should also be possible to construct semi-classical Green functions in case the pair $(\Lambda,\Lambda_+)$
is no longer intersecting cleanly, but glancing. We then need second-microlocalization, and introducing
so called 2-phases, see e.g. [LaWi] in the standard (polyhomogeneous) calculus. 
\bigskip
\noindent {\it Acknowledgements}: This work was supported by Grant 
PRC No. 1556 CNRS-RFBR 2017-2019 
``Multi-dimensional semi-classical problems of 
Condensed Matter Physics and Quantum Mechanics'', and RFBR Grant No. 17-51-150006.

\bigskip
\centerline {\bf References}
\medskip
\noindent [AnDoNaRo1] A.Anikin, S.Dobrokhotov, V.Nazaikinskii, M.Rouleux. Maslov's canonical operator on a pair of 
Lagrangian manifolds and asymptotic solutions of stationary equations with localized right-hand sides.
Doklady Akad. Nauk, Vol. 76, No1, p.1-5, 2017.

\noindent [AnDoNaRo2] A.Anikin, S.Dobrokhotov, V.Nazaikinski, M.Rouleux.
Asymptotics of Green function for the linear waves equations in a domain with a non-uniform bottom.
Proceedings ``Days of Diffraction 2017'', Saint-Petersburg, IEEE p.18-23.

\noindent [Ba] V.M.Babich, On the short-wave asymptotic behavior of Green's function
for the Helmholtz equation, Math. Sb. 65(107) (4), p.576-630, 1964.

\noindent [Bad] N.Baddour, Operational and convolution properties of two-dimensional Fourier transforms in polar coordinates,
J. Opt. Soc. Am. A, Vol.26, p.1767-1777, 2009.

\noindent [CdV] Y.Colin de Verdi\`ere. M\'ethodes semi-classiques et th\'eorie spectrale. 
https://www-fourier.ujf-grenoble.fr/~ycolver/ All-Articles/93b.pdf

\noindent [DoNa] S.Dobrokhotov, V.Nazaikinskii. Punctured Lagrangian manifolds and asymptotic solutions of the
linear water-wave equations with localized initial solutions.
Math. Notes, 101, No.6, p.130-137, 2017.

\noindent [DoMaNa] S.Dobrokhotov, G. Makrakis, V.Nazaikinskii. Fourier Integrals and a new representaion of Maslov's
canonical operator near caustics. Amer. Math. Soc. Trans. Vol. 233, p.95-115, 2014.

\noindent [DNS] S.Dobrokhotov, V.Nazaikinskii, A.Shafarevich. New integral representations of Maslov canonical operators
in singular charts. Izvestiya Math. 81(2), p.286-328, 2017.

\noindent [DoZh] S.Dobrokhotov, P.Zhevandrov. Asymptotic expansions and the Maslov canonical 
operator in the linear theory of water-waves I. Main constructions and equations for surface 
gravity waves. Russian J. Math. Phys. Vol.10 No.1, p.1-31, 2003.

\noindent [Du] J.J.Duistermaat. Oscillatory integrals, Lagrangian immersions and unfolding of singularities, 
Commun. Pure Applied Math., Vol.27, p.207-281, 1974.

\noindent [HeSj] B.Helffer, J.Sj\"ostrand. Multiple wells in the semi-classical limit I. Comm. Part. Diff. Eqn. 9(4) p.337-408, 1984.

\noindent [H\"o] L.H\"ormander. {\it The Analysis of Linear Partial Differential Operators I,IV}. Springer, 1985.

\noindent [Iv] V.Ivrii. {\it Microlocal Analysis and Precise Spectral Asymptotics}. Springer-Verlag, Berlin, 1998.

\noindent [Ke] J.B. Keller, Geometrical Theory of Diffraction,  J. Opt. Soc. Am. 52, p.116-130, 1962.

\noindent [Ki] A.P.Kiselev. Localized light waves: Paraxial and exact solutions of the wave equation (a review). Optics
and Spectroscopy, Vol.102 (4), p.603-622, 2007. 

\noindent [Ku] V.Kucherenko. Quasi-classical asymptotics of a point source function for the stationary Schr\"o- dinger equation.
Teor. i Mat. Fiz. Vol.1, No.3, p.384-406. Transl. Consultants Bureau, 1970. 

\noindent [LaWi] P.Laubin, B.Willems. Distributions associated to a 2-microlocal pair of Lagrangian manifolds.
Comm. Part. Diff. Eq. 19(9\& 10), p.1581-1610, 1994.

\noindent [M] V.P.Maslov. {\it Th\'eorie des Perturbations et M\'ethodes Asymptotiques}. Dunod, Paris, 1972.

\noindent [MelUh] R.Melrose and G.Uhlmann,
Lagrangian intersection and the Cauchy problem,
Comm. Pure Appl. Math. 32 (4), p.483-519, 1979.

\end